\pdfoutput=1
\documentclass[prl,twocolumn,superscriptaddress,preprintnumbers,nofootinbib]{revtex4}
\usepackage{graphicx}
\usepackage{epsfig}
\usepackage{bm}
\usepackage{latexsym,amssymb,amsmath,amsfonts,amssymb,txfonts,pxfonts,wasysym,float}
\usepackage{mathrsfs}
\usepackage{color}
\usepackage{lettrine}
\usepackage{lipsum}
\usepackage{enumitem}

\newcommand{\postscript}[2]{\setlength{\epsfxsize}{#2\hsize}
   \centerline{\epsfbox{#1}}}

 \def\be{\begin{equation}}
\def\ee{\end{equation}}
\def\ba{\begin{eqnarray}}
\def\ea{\end{eqnarray}}
\usepackage{pdfsync}

\usepackage[usenames,dvipsnames]{xcolor}
\definecolor{orange}{cmyk}{0,0.5,1,0}
\definecolor{rossoCP3}{cmyk}{0,.88,.77,.40}
\definecolor{graa}{rgb}{0.8,0.8,0.8}
\definecolor{blaa}{rgb}{0.2,0.2,0.6}

\begin{document}

\preprint{MPP-2021-68}
\preprint{LMU-ASC 12/21}

\title{\color{rossoCP3} Leptophilic $\bm{U(1)}$ Massive Vector Bosons from Large Extra Dimensions}

\author{Luis A. Anchordoqui}

\affiliation{Department of Physics and Astronomy,  Lehman College, City University of
  New York, NY 10468, USA}

\affiliation{Department of Physics,
 Graduate Center, City University
  of New York,  NY 10016, USA}

\affiliation{Department of Astrophysics, 
 American Museum of Natural History, NY
 10024, USA}

\author{Ignatios Antoniadis}

\affiliation{Laboratoire de Physique Th\'eorique et Hautes \'Energies - LPTHE
Sorbonne Universit\'e, CNRS, 4 Place Jussieu, 75005 Paris, France}

\affiliation{Institute for Theoretical Physics, KU Leuven, Celestijnenlaan 200D, B-3001 Leuven, Belgium}

\author{Xing~Huang}
\affiliation{Institute of Modern Physics, Northwest University, Xi'an
  710069, China}
\affiliation{NSFC-SPTP Peng Huanwu Center for Fundamental Theory,
  Xi’an 710127, China}
\affiliation{Shaanxi Key Laboratory for Theoretical Physics Frontiers,
  Xi'an 710069, China}

\author{Dieter L\"ust}

\affiliation{Max--Planck--Institut f\"ur Physik, 
 Werner--Heisenberg--Institut,
80805 M\"unchen, Germany
}

\affiliation{Arnold Sommerfeld Center for Theoretical Physics 
Ludwig-Maximilians-Universit\"at M\"unchen,
80333 M\"unchen, Germany
}

\author{Tomasz R. Taylor}

\affiliation{Department of Physics,
  Northeastern University, Boston, MA 02115, USA}

\begin{abstract}
  \vskip 2mm \noindent We demonstrate that the discrepancy between the
  anomalous magnetic moment measured at BNL and Fermilab and the
  Standard Model prediction could be explained within the context of
  low-scale gravity and large extra-dimensions. The dominant
  contribution to $(g-2)_\mu$ originates in Kaluza-Klein (KK)
  excitations (of the lepton gauge boson) which do not mix with quarks (to lowest order) and therefore can be quite light
  avoiding LHC constraints. We show that the KK contribution to
  $(g-2)_\mu$ is universal  with the string scale entering as an effective cutoff. The KK tower provides a unequivocal
  distinctive signal which will be within reach of the future muon
  smasher.
\end{abstract}

\maketitle

Low scale gravity and large extra dimensions offer a genuine solution
to the gauge hierarchy problem~\cite{ArkaniHamed:1998rs,Antoniadis:1998ig}. Within these models
one has to address the problem of baryon $B$ and lepton $L$ number violation
by higher dimensional operators suppressed only by the low string
scale $M_s$. Intersecting D-brane models offer a way out by gauging these
symmetries~\cite{Antoniadis:2000ena,Blumenhagen:2000wh,Antoniadis:2002qm,Blumenhagen:2005mu,Blumenhagen:2006ci}. Since the $B$ and $L$ gauge bosons are anomalous they 
gain masses through a generalization of the Green-Schwarz (GS) anomaly cancellation~\cite{Sagnotti:1992qw,Ibanez:1998qp,Poppitz:1998dj,Ibanez:1999it} giving rise to
perturbative global symmetries broken only by non-perturbative effects
that are suppressed exponentially by the string/gauge coupling. The
resulting gauge bosons form in general linear combinations of the various abelian gauge factors orthogonal to the hypercharge combination, that couple
to both quark and leptons. However, the Kaluza-Klein (KK) excitations
do not mix (to lowest order) and thus those of $L$ couple only to
leptons. Such modes can be quite light because LHC constraints are
weak but can provide a sizable contribution to the anomalous magnetic moment of the
muon $a_\mu = (g-2)_\mu/2$.

TeV-scale D-brane string compactifications could then provide an
innovative framework to explain the extant tension between the
Standard Model (SM) prediction of $a_\mu$ and experiment. Very
recently, the Muon $g-2$ Experiment at Fermilab reported a measurement
reading
$a_\mu^{\rm FNAL} = 116592040(54)\times10^{-11}$~\cite{Abi:2021gix},
which is larger than the SM prediction $a_\mu^{\rm
  SM}=116591810(43)\times10^{-11}$ in which contributions from QED, QCD, and electroweak interactions are taken into account with highest
precision~\cite{Aoyama:2020ynm}. This leads to
$a_\mu^{\rm FNAL} - a_\mu^{\rm SM} = (230 \pm 69) \times 10^{-11}$,
which corresponds to a $3.2\sigma$ discrepancy. Because the Fermilab
observation is
compatible with the long-standing discrepancy from the E821 experiment
at BNL~\cite{Bennett:2006fi}, the overall deviation from the SM central value, 
\begin{equation}
\Delta a_\mu^{\rm exp} \equiv a_\mu^{\rm FNAL + BNL} - a_\mu^{\rm SM} = (251
\pm 59) \times 10^{-11} \,,
\label{deltaamuexp}
\end{equation}
strengthens the significance to
$4.2\sigma$~\cite{Abi:2021gix}.\footnote{We note in passing that the SM prediction estimated by the latest
  lattice QCD calculations, $a_\mu^{\rm
    SM,lattice}=11659195163(58)\times10^{-11}$, has a larger
  uncertainty and brings the prediction closer to the experimental value, $a^{\rm FNAL + BNL}_\mu - a^{\rm SM,lattice}_\mu
  =   109(71)\times10^{-11}$, yielding only a $1.6~\sigma$
  effect~\cite{Borsanyi:2020mff}.} Even though the discrepancy is not statistical significant yet, it is
interesting to entertain the possibility that it corresponds to a real
signal of new physics. In this Letter we calculate the massive vector boson contribution
  to $g-2$ from KK excitations of $L$ and we show that it is {\it universal}
  and can accommodate
  the $\Delta a_\mu^{\rm exp}$ discrepancy of (\ref{deltaamuexp}).

\begin{figure}[tbp] 
  \postscript{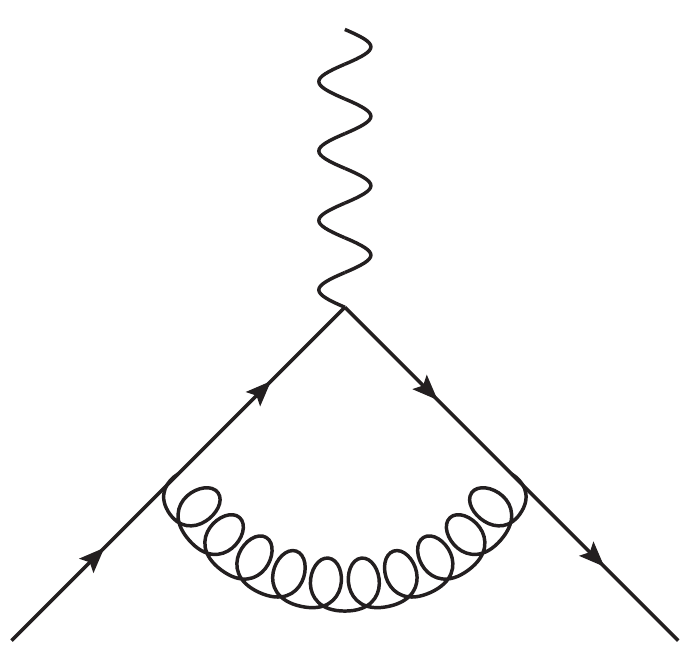}{0.8}
\caption{KK gauge boson (double wavy line) contribution to muon's anomalous magnetic moment.
\label{g2:fig1}}
\end{figure}

At the leading  order in the $U(1)_L$ coupling constant $g_L$, the contribution of a massive vector boson to lepton's $g-2$ originates from the vertex correction shown in Fig.~\ref{g2:fig1}. 
Note that KK momentum is not conserved in lepton gauge boson vertices
since leptons are localized in brane
intersections. Figure~\ref{g2:fig1} shows the same diagram that yields
the famous $\alpha/\pi$ in QED, but with the virtual photon replaced
by a massive vector boson. The fastest way to compute it is to use the massive propagator in unitary gauge, \be
D^{\mu\nu}(k)=\frac{-i}{k^2-M^2}\bigg(g^{\mu\nu}-\frac{k^\mu k^\mu}{M^2}\bigg),\label{prop}\ee
and follow a textbook, for example Ref.\cite{sch}. It is easy to see that the longitudinal part of the propagator (second term in Eq.(\ref{prop})) does not contribute to the magnetic moment. The only difference between the first term and the photon propagator in Feynman gauge is $M^2$ in the denominator which leads to a slight modification of the integral over Feynman parameters. In the limit of $M\gg m$, one obtains
\be\Delta a_\mu=\frac {(g-2)_\mu} 2= \frac{1}{3}{\alpha_L\over \pi}{m^2\over
  M^2}\ ,
\label{Delta1}
\ee
where we neglected terms of order $(m/M)^4$ and $m$ is the muon mass. Note that this is a {\em
  positive} correction that brings $a_\mu$ closer to experimental
data.

The masses of KK gauge bosons are labelled by integer vectors $\vec n$, with $M^2(\vec n)=|\vec n|^2M^2$, where $M$ is the compactification scale. For $M\ll M_s$ the couplings $\alpha_L(\vec n)$ depend very mildly on $\vec n$  when $|\vec n|$ is small \cite{Antoniadis:1993jp}. They are approximately 1 until $|\vec n|\approx M_s/M$ and then exponentially suppressed when $|\vec n|\gg M_s/M$. They are given by a Gaussian form $\alpha_L(\vec n)=\delta^{-{{\vec n}^2 M^2/M_s^2}}$ with $\delta<1$ a model dependent constant.
In the case of one extra dimension
\be\Delta a_\mu= \sum_n\frac{1}{3}{\alpha_L(n)\over \pi}{m^2\over n^2M^2}\approx
{\alpha_L\pi\over 18}{m^2\over M^2}\ .\ee
In the case of two extra dimensions, the exponential suppression of $\alpha_L(\vec n)$ at large 
$|\vec n|$ is crucial for regulating the logarithmic divergence of the sum:
\be\Delta a_\mu= \sum_{\vec n}\frac{1}{3}{\alpha_L(\vec n)\over \pi}{m^2\over |\vec n|^2M^2}\approx
{2\alpha_L\over 3}{m^2\over M^2} \ \ln\left(\frac{M_s}{M}\right) .\ee
Here, $\alpha_L$ is the coupling of the lightest KK excitation.
\begin{figure}[tbp] 
  \postscript{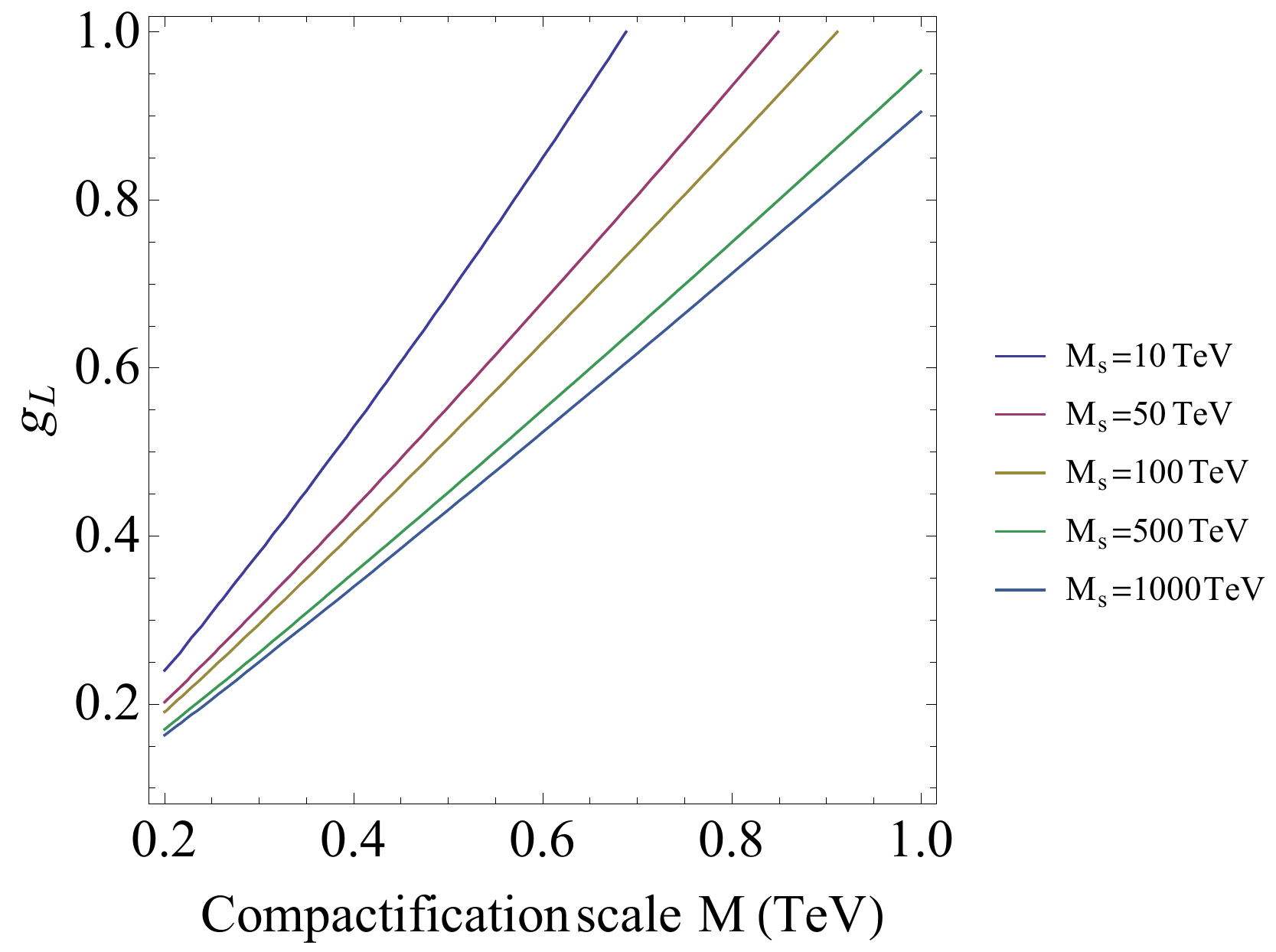}{0.99}
\caption{Contours of constant $\Delta a_\mu^{\rm exp}$ for different
  values of $M_s$.
\label{g2:fig2}}
\end{figure}

To develop some sense for the orders of magnitude involved, we recall that
direct production at LEP provides the best bound on KK couplings and masses.
The agreement between the LEP-II measurements and the SM predictions
implies that either
$g_L \alt 10^{-2}$, or else $M > 209~{\rm GeV}$, the maximum energy of
LEP-II~\cite{Appelquist:2002mw}. In Fig.~\ref{g2:fig2} we show contours
of constant $\Delta a_\mu^{\rm exp}$ in the $g_L-M$ plane for
different values of the string scale. We see that there is a large
range of masses and couplings that can accommodate the Fermilab
result.  A point worth noting at this juncture is that the KK 
contribution to $(g-2)_\mu$ is universal, with $M_s$ entering as
an effective cutoff.

There are two different classes of D-brane constructs that can realize
the tower of KK modes. On the one hand, we can envision that $L$ is part of the
hypercharge (thus its gauge coupling $\alpha_L$ cannot be very small). 
One can then try to use one of the orthogonal to the hypercharge combinations for explaining the
$(g-2)_\mu$ discrepancy and make it leptophilic to avoid the LHC bounds. It turns out that this 
cannot be done because the corresponding $U(1)$ gauge coupling becomes strong.
Indeed, the 4 stack model thoroughly
analyzed in~\cite{Anchordoqui:2021llp}, with gauge
group $U(3)_a \times Sp(1)_b \times U(1)_c \times U(1)_d$, typifies
this class.  Contact with gauge structures at TeV energies is achieved by a field rotation to couple diagonally to
hypercharge $Y$. Two of the Euler angles ($\psi,\theta,\phi$) are determined by this rotation. The gauge couplings are related to $g_Y$ by
\be
\frac{1}{(6 g'_a)^2} + \frac{1}{(2 g'_c)^2}+\frac{1}{(2 g'_d)^2}=\frac{1}{g_Y^2} \, ,
\label{conharbinger}
\ee
and the relation for $U(N)$ unification, $g'_N = g_N / \sqrt{2N}$, holds only at $M_s$ because the $U(1)$ couplings ($g'_a,$ $g'_c$,
$g'_d$) run differently from the non-abelian $SU(3)$ ($g_a$) and
$SU(2) \equiv Sp(1)$ 
($g_b$)~\cite{Anchordoqui:2011eg}. The zero-mode of the anomalous $U(1)$,
hereafter $Z'$, gains a mass via the GS mechanism by absorbing an axionic field from the  R-R (Ramond) closed string sector. To get as much contribution to $a_\mu$ as possible without violating
the LHC bounds~\cite{Sirunyan:2019vgj,Sirunyan:2021khd}, it is natural to consider a leptophilic (in our case
meaning large $g_L \equiv g'_c$) $Z'$~\cite{Buras:2021btx}. Next, we compare with the LHC data considering
the resonant production cross section of $\sigma(pp\to Z' \to \ell
\ell)$. Under the narrow width approximation, the cross section can be written in the form of $c_u w_u + c_d w_d$, where $w_u, w_d$ are given by model-independent parton distribution functions~\cite{Accomando:2010fz}. The coupling of $Z'$ with up and down quarks (assuming same coupling to three families)
are encoded in $c_u,c_d$. More precisely, for a generic coupling between $Z'$ and fermion~$f$\begin{equation}
Z^\prime_{\mu}\gamma^{\mu}(\bar f_L\epsilon_L^f f_L +\bar f_R \epsilon_R^f
f_R)\,,
\end{equation}
the coefficients $c_u$ and $c_d$ take the following form
\begin{equation}
c_{f}=({\epsilon_L^f}^2+{\epsilon_R^f}^2)\textrm{Br}(\ell^+\ell^-)\,.
\end{equation}
We compute the branching faction $\textrm{Br}(\ell^+\ell^-)$ by including
only the decay channels to leptons and quarks. The total decay rate is given
by
\begin{equation}
\Gamma_{Z'}=
\frac{1}{24\pi}M_{Z'}\left[
9\sum_{q=u,d} ({\epsilon_L^q}^2+{\epsilon_R^q}^2) +3\sum_{\ell=e,\nu} ({\epsilon_L^\ell}^2+{\epsilon_R^\ell}^2)
\right].
\end{equation}
Due to the constraint
(\ref{conharbinger}),  there are two free parameters (for a given
string scale $M_s$): $\phi$ and $g'_d (M_s)$. Setting the mass of $Z'$ to 2~TeV, we then search
over the parameter space to get the smallest possible values of $c_u,
c_d$. For simplicity, the combination of $\sqrt{c_u^2+c_d^2}$ is
considered. We find that the optimal value of $\phi$ generally
suppresses the couplings to left-handed quarks and the remaining
couplings to the right-handed quarks are controlled by $g'_d$. In the
best case scenario, $g'_c(M_s)$ is set to $2\pi$ at $M_s$ (with $10
\alt M_s/{\rm TeV} \alt 10^3$), the corresponding cross section (or rather $\sqrt{c_u^2+c_d^2}
\sim 8.4\times 10^{-5}$) is roughly 2 percent of that given by the
sequential standard model boson $Z'_{\textrm{SSM}}$~\cite{Altarelli:1989ff}, saturating the
LHC limit~\cite{Sirunyan:2021khd}. We note that the
branching fraction to leptons is close to $1$ due to the small
coupling to quarks. The signal can be further reduced by including
other decay channels. Moreover, the largest possible $g'_c(M_s)$ also
gives the most contribution to $a_\mu$. Such a $Z'$ boson gives $a_\mu
= 9.9 \times 10^{-11}$~\cite{Anchordoqui:2021llp},  which
  is much smaller than the pre-LHC estimate of Ref.~\cite{Kiritsis:2002aj} and it is not enough to explain the observed
discrepancy. The second anomalous $U(1)$ should be much heavier to avoid the LHC bound
and its contribution to $a_\mu$ is negligible. To accommodate the Fermilab data
one can advocate the violation of lepton flavor
universality~\cite{Buras:2021btx}.  Alternatively, as we have shown in Fig.~\ref{g2:fig2}, the Fermilab/BNL data can be interpreted as evidence for {\it massive vector boson contributions
  to $g-2$ from KK excitations of the $U(1)_c$.}
Note that in contrary to the gauge boson 0-mode which acquires a mass from the anomaly, the masses of KK modes originate from the internal component(s) of the higher dimensional gauge field.

On the other hand, we can envision that $L$ is not part of the
hypercharge. If this were the case, the KK tower and even its (anomalous) zero-mode would be completely unconstrained. The generic features of the D-brane constructs (with more than 4 stacks
of D-branes) that can realize this class of models can be summarized as follows:
\begin{itemize}[noitemsep,topsep=0pt]
\item the lepton doublet should lie on the intersection of the weak
  $U(2)_w$ and $U(1)_L$, so that the Abelian charge $Q_L$ participates in the hypercharge $Y$ but not $L$;
\item the lepton $l^c$ should lie on an intersection of a $U(1)$ that participates in $Y$ and $U(1)_L$ so that has opposite lepton charge from $l$;
\item the quarks should not see $U(1)_L$;
\item $U(1)_L$ can even be in the bulk (or part of it) with no important accelerator constraints.
\end{itemize}

In summary, we have shown that the exchange of KK excitations of
the $L$ (lepton number) gauge boson can provide a dominant contribution to
$(g-2)_\mu$ and explain the $\Delta a_\mu^{\rm ex}$ discrepancy
reported by BNL and Fermilab.  In the case of two extra dimensions, the
summation of KK modes gives an additional factor of ${\cal O} (10)$
change in the prediction for $\Delta a_\mu$ compared to that of a
single $Z'$-gauge boson, and this is pivotal to avoid the violation of
lepton flavor universality in accommodating the data. The KK tower, which will be within reach of the future muon
  smasher~\cite{Ali:2021xlw}, may become the smoking gun of
  low-scale gravity models and large extra dimensions.\\

\acknowledgments{The work of L.A.A. is supported by the by the 
  U.S. National Science Foundation (NSF Grant PHY-2112527). The research of I.A. was partially performed as
  International professor of the Francqui Foundation, Belgium. The
  work of X.H. is supported by the NSFC Grant No. 12047502. The
  work of D.L. is supported by the Origins Excellence Cluster. The
  work of T.R.T is supported by NSF under Grant Number PHY-1913328.
  Any opinions, findings, and conclusions or recommendations expressed
  in this material are those of the authors and do not necessarily
  reflect the views of the NSF.}

\end{document}